\documentclass[aps,prl,twocolumn,superscriptaddress,nofootinbib,nobalancelastpage,showpacs,nobibnotes]{revtex4}
\usepackage{epsfig}
\begin{document}
\title{Converting a series in $\lambda$ to a series in $\lambda^{-1}$}
\author{Andrew A. Rawlinson}
\email{a.rawlinson@physics.unimelb.edu.au}
\affiliation{School of Physics, The University of Melbourne,  Victoria 3010, Australia}

\begin{abstract}
We introduce a transformation for converting a series in a parameter, $\lambda$, to a series in the inverse of the parameter $\lambda^{-1}$. By applying the transform on simple examples, it becomes apparent that there exist relations between convergent and divergent series, and also between large- and small-coupling expansions. The method is also applied to the divergent series expansion of Euler-Heisenberg-Schwinger result for the one-loop effective action for constant background magnetic (or electric) field. The transform may help us gain some insight about the nature of both divergent (Borel or non-Borel summable series) and convergent series and their relationship, and how both could be used for analytical and numerical calculations. 
\end{abstract}

\pacs{11.15Bt,  11.10Jj, 11.15Me, 0.23.Lt}

\maketitle

\section{Introduction}

Much of our understanding of physics and mathematics relies on our ability not only to find relationships between various quantities through equations (such as differential equations), but also their solutions. In many cases one resorts to assuming that the solution can be expressed as a series expansion $\sum c(n) \lambda^n$ in some parameter $\lambda$. Depending on the nature of the problem at hand, a first attempt at a series expansion might yield a convergent series, whereas in other situations, a divergent series. One finds in many instances that a divergent series arises, e.g. in perturbative expansions in field theories~\cite{dyson,hurst,thirring}

Convergent series are generally well behaved. Divergent series, on the other hand, require more careful attention to elucidate their meaning. In many scenarios, the information one obtains, such as an expression for the coefficient function $c(n)$, is very useful and various schemes has been proposed that allow to one assign values to divergent series. 
Many analytical and numerical methods have been put forward, e.g. Borel summation, Pad\'e and Borel-Pad\'e approximants, large order perturbation theory (see~\cite{LeGandZJ} and references therein), variational perturbation theory (\cite{kleinertPhi4} and refs. therein), sequence transformations (\cite{weniger} and refs. therein).

Another, but little known, numerical procedure is that of Mellin-Barnes regularisation originally proposed by Kowalenko~{\it et al}~\cite{kowalenko,kowANDtau,kowANDraw}, which can be used to obtain finite numerical values of a divergent (or convergent) series, for certain values of complex $\lambda$. This method is an exact procedure in the sense that no approximations are made, and it yields the same numerical values as those obtained by Borel summation for Borel summable series.

Further investigation of the numerical Mellin-Barnes regularisation method, particularly with coefficient functions $c(n)$ containing $\Gamma$-functions, and extending $\lambda$ to the entire complex plane (i.e. including non-Borel summable series), revealed an analytical relationship between series expansions in $\lambda$ to those in $\lambda^{-1}$. This is in contrast to the basic Borel transformation, in which case, before and after the transform one still has a series in $\lambda$.

With this new insight we show how one can transform a series in $\sum c(n) \lambda^n$ to a series in $\sum d(n) \lambda^{-n}$. It then becomes apparent that the transform makes no distinction between Borel (alternating) and non-Borel (non-alternating) summable series. In fact, in some cases it converts one to the other, and also converts a divergent series to a convergent series. As we shall see, depending on the form of the coefficient function $c(n)$, a single series in $\lambda$ can yield more than one series in $\lambda^{-1}$.

Many functions have both convergent and divergent series, and integral, representations. Our ability to understand the solution to a problem is only as good as our ability to extract information about the coefficient function $c(n)$. An advantage of the proposed transformation is that given $c(n)$, we can look at both the series in $\lambda$ and in $\lambda^{-1}$. If one has the exact expression for $c(n)$, be it as a convergent or divergent series, then it can be regarded as the solution to the problem. If we only have an approximate expression for $c(n)$, it is hoped that converting $\sum c(n)\lambda^n$ to $\sum d(n) \lambda^{-n}$ might allow us to gain more understanding of the solution to the problem. 

\section{The Transform}
Consider a sum of a finite number of terms
\begin{equation}
S(x,N)=\sum_{n=0}^N c(n) x^n
\end{equation}
for any (meromorphic) coefficient function $c(n)$. Using Cauchy's integral theorem, we can express this as
\begin{equation}
S(x,N)=\oint_{C(N)}\,dt\, {c(t)\, x^t \over \exp(2\pi i t)-1}\equiv \oint_{C(N)}
\label{theINT}
\end{equation}
where we have introduced a short hand notation for the integral. It is assumed that the only poles within the contour $C(N)$ are the simple poles (of residue $1/(2\pi i)$) of $1/(\exp(2\pi i t)-1)$ at $t=0,1,2,\ldots,N$. (Of course, if there are poles due to $c(t)$, their residues are included).

\begin{figure}[ht]
\begin{center}
\epsfig{file=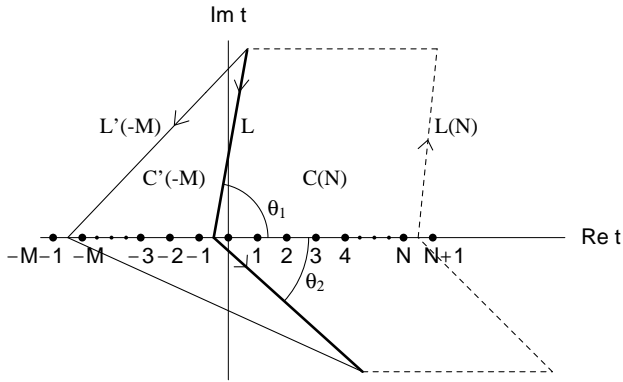,width=3.25in}
\caption{\label{figure}
The complex $t$-plane showing the regions enclosed by the contours $C(N)=L(N)+L$ and $C'(-M)=L'(-M)-L$.}
\end{center}
\end{figure}

If we express the closed contour $C(N)=L(N)+L$ where $L$ intersects the $Re\,t$ axis somewhere at $-1<a<0$, and $L(N)$ intersects the $Re\,t$ axis between $N$ and $N+1$ as shown in Fig.~\ref{figure}, the nature of $c(t) x^t/(\exp(2\pi i t)-1)$ in the complex $t$-plane may allow one to adjust the angles $0<\theta_1<\pi$ and $-\pi<\theta_2<0$ of $L$ so that 
\begin{eqnarray}
& &\lim_{R\rightarrow\infty}\, {c(t)\, x^t \over \exp(2\pi i t)-1}\longrightarrow 0 \nonumber \\
{\rm for\,\,} t&=& a\exp(i\pi)+R\exp(i\theta_1) \nonumber \\ 
{\rm and\,\,} t&=&a\exp(i\pi)+R\exp(i\theta_2)
\end{eqnarray}
where we note the angles are centred at $t=(-a,0)=a\exp(i\pi)$, rather than at the origin $t=(0,0)$. If these limits can be achieved, the integral along $L$ is independent of $N$. This is possible if $c(t)$ contains $\Gamma$-functions. See Fig.~\ref{zABm25} for an example, where the coefficient function is given by~(\ref{classAHOdsfinitesum}). Note that the choice of contours $L$ and $L(N)$ will depend on both $x$ and $c(t)$, since for a given $c(t)$ it is possible one will need a different contour for $x=2$, say, compared to $x=-2$.

Using the short hand notation in~(\ref{theINT}), we can express the loop integration in~(\ref{theINT}) as
\begin{eqnarray}
\oint_{C(N)}&=&\int_{L(N)}+\int_{L} \nonumber \\ 
&=& 2\pi i \sum {\it Res(C(N))} = \sum_{n=0}^N c(n) x^n
\label{resRHS}
\end{eqnarray}
where $\sum {\it Res(C(N))}$ denotes the sum of the residues of the poles within $C(N)$, where the last equality is based on the assumption that $c(t)$ has no poles within $C(N)$.

Now consider another contour $C'(-M)=L'(-M)-L$, where $L'(-M)$ intersects the $Re\,t$ axis somewhere between $-M-1$ and $-M$, as shown in Fig.~\ref{figure}, and we note the sign in front of $L$ is changed because of the counterclockwise convention of Cauchy's intergral theorem. One finds that
\begin{equation}
\oint_{C'(-M)}=\int_{L'(-M)}-\int_L =2\pi i  \sum {\it Res(C'(-M))}
\label{resLHS}
\end{equation}
where $\sum {\it Res(C'(-M))}$ is the sum of the residues of the poles within $C'(-M)$. Thus, combining~(\ref{resRHS}) and~(\ref{resLHS}), we find that
\begin{eqnarray}
\int_L&=& \sum_{n=0}^N c(n) x^n-\int_{L(N)} \nonumber \\ 
&=& -2\pi i  \sum {\it Res(C'(-M))}+\int_{L'(-M)}.
\label{mainresult}
\end{eqnarray}
This is the main result of the paper, and is applicable to cases where $c(n)$ represents either a convergent or divergent series. One can use this result for analytical and numerical studies of a series. In numerical studies, if the contour $L$ in Fig.~\ref{figure} is chosen with $\theta_1=\pi/2$ and $\theta_2=-\pi/2$, the first line of~(\ref{mainresult}) gives the Mellin-Barnes regularisation~\cite{kowalenko,kowANDtau,kowANDraw}.

In general, one often finds that if $\sum_{n=0}^N c(n) x^n$ is a divergent series, then $ \sum {\it Res(C'(-M))}$ is a convergent series, and the nature of the integrand in~(\ref{resLHS}) is such that as $M\rightarrow\infty$ then $\int_{L'(-M)}\rightarrow 0$, so 
\begin{equation}
\sum_{n=0}^N c(n) x^n-\int_{L(N)} = -2\pi i  \sum {\it Res(C'(-\infty))} = \int_L.
\end{equation}
Therefore we find that $-2\pi i  \sum {\it Res(C'(-\infty))} $ is finite and can be regarded as the convergent series representation of $\sum_{n=0}^N c(n) x^n$, and also of $ \int_L$. Observe that since $\sum_{n=0}^N c(n) x^n$ is a divergent series, it grows with $N$, but so does $\int_{L(N)}$, in such a way that $\sum_{n=0}^N c(n) x^n-\int_{L(N)}=\int_L$ is finite and independent of $N$, i.e.
\begin{eqnarray}
\int_L &=& \sum_{n=0}^N c(n) x^n-\int_{L(N)} \nonumber \\
&=& \sum_{n=0}^{N'}c(n) x^n-\int_{L(N')}.
\end{eqnarray}

\section{Some examples}
Let consider an example with a series having a coefficient function of $c(t)= \Gamma(t+\alpha)$ where, without loss of generality, we take $\alpha>0$ and a non-integer. Subsituting this into~(\ref{mainresult}) we see that the only poles within $C(N)$ are simple poles at $t=0,1,2,\ldots$ due to $1/(\exp(2\pi i t)-1)$, and their residues sum up to yield the divergent series
\begin{equation}
2\pi i\sum Res(C(N))=\sum_{n=0}^N\Gamma(n+\alpha) x^n. 
\label{alphaDS}
\end{equation}

Within $C'(-M)$ we find two sets of simple poles. One set at $t=-1,-2,-3,\ldots$ from  $1/(\exp(2\pi i t)-1)$ where the residue of the pole at $t=-m$ is
\begin{equation}
x^{-m}\Gamma(\alpha-m)={\pi\over (-x)^m \sin(\alpha\pi) \Gamma(m+\alpha-1)}
\label{alphaCS1}
\end{equation}
where we have used the $\Gamma$-function reflection formula, $\Gamma(x)\Gamma(1-x)=\pi/\sin(\pi x)$. Summing these residues, we see that it is a convergent series and therefore we can let $M\rightarrow\infty$ :--
\begin{eqnarray}
& &\sum_{m=1}^\infty{\pi\over (-x)^m \sin(\alpha\pi) \Gamma(m+\alpha-1)}\nonumber \\
& &\hspace{5mm}= {\pi e^{-1/x}\over (-x)^\alpha\sin(\alpha\pi)}\left(1-{\Gamma(1-\alpha,-1/x)\over\Gamma(1-\alpha)}\right)
\label{alpha1series}
\end{eqnarray}
where $\Gamma(y,z)$ is the incomplete $\Gamma$-function. This series, arising from the poles of $1/(\exp(2\pi i t)-1)$, can be regarded as the counterpart of~(\ref{alphaDS}) where we take the sum over negative, instead of positive, integers.

The other set of poles at $t=-\alpha,-\alpha-1,-\alpha-2,\ldots$ are from $\Gamma(t+\alpha)$, and the residue of the pole at $t=-\alpha-m$ is
\begin{equation}
{-\pi\over (-x)^{\alpha+m} \sin(\alpha\pi) \Gamma(m+1)}
\label{alphaCS2}
\end{equation}
and summing up these residues gives a convergent series as well :--
\begin{equation}
\sum_{m=0}^\infty{-\pi\over (-x)^{\alpha+m} \sin(\alpha\pi) \Gamma(m+1)} =
 {-\pi e^{-1/x}\over (-x)^\alpha\sin(\alpha\pi)}
\end{equation}
which will cancel the first term in the brackets in~(\ref{alpha1series}). This series arises from the poles in the coefficient function $c(t)=\Gamma(t+\alpha)$.
Therefore, the total sum of residues within $C'(-\infty)$ is
\begin{equation}
2\pi i \sum Res(C'(-\infty))
={-e^{-1/x}\Gamma(\alpha)\Gamma(1-\alpha,-1/x)\over (-x)^\alpha}
\label{alphaCS}
\end{equation}
where we have used the $\Gamma$-function reflection formula again. We thus see the relation of the divergent series~(\ref{alphaDS}) and the convergent series~(\ref{alphaCS}) via~(\ref{mainresult}). 

If we now consider $\alpha$ to be an integer in~(\ref{alphaDS}), say, $\alpha=1$, we get the so-called `paradigm divegerent series' :--
\begin{equation}
\sum_{n=0}^N \Gamma(n+1) x^n.
\label{paraDS}
\end{equation}
Inserting this into~(\ref{mainresult}), we have simple poles at $t=0,1,2,\ldots$ within $C(N)$, and the sum of their residues yields the series~(\ref{paraDS}). Since $\Gamma(t+1)$ has poles at $t=-1,-2,-3,\ldots$, we now have double poles within $C'(-M)$, where the residue of the pole at $t=-m$ is
\begin{equation}
{-(\log(-x)+\psi(m))\over (-x)^m\Gamma(m)}
\label{paraLOG}
\end{equation}
where $\psi(n)$ is the digamma function. Since the sum of these residues is a convergent series, we can take $M\rightarrow\infty$, and obtain
\begin{eqnarray}
2\pi i\sum Res(C'(-\infty))&=&\sum_{m=1}^\infty {-(\log(-x)+\psi(m))\over (-x)^m\Gamma(m)}\nonumber \\
&=&{e^{-1/x} \Gamma(0,-1/x)\over x}
\label{paraANS}
\end{eqnarray}
Note that this is the same as~(\ref{alphaCS}) with $\alpha=1$. Notice the presence of $\log(-x)$ in each residue term in~(\ref{paraLOG}), which can pulled out of the summation in~(\ref{paraANS}). However, the summation of the $\psi(m)$ terms yields a $\log(-1/x)$ term, such that it precisely cancels the $\log(-x)$, resulting in the $\log(-x)$-free expression in~(\ref{paraANS}). This is an indication whereby if one finds logarithmic terms in a series expansion, that upon summation of the series such terms might vanish.

Thus we see some of main features of~(\ref{mainresult}) :-- a series in $x^n$ is converted to a series in $(-1/x)^n$, i.e. a non-alternating series is converted to an alternating series in the inverse of the expansion parameter. If double, or higher order, poles are encountered, derivatives (wrt $t$) of the coefficient function, $c(t)$, and power term, $x^t$, yield derivatives of $\Gamma$-functions and $\log(x)$ terms respectively. We note that if $x<0$, then~(\ref{paraANS}) is real, whereas if $x>0$, a non-zero imaginary part arises of magnitude $-(\pi/x) \exp(-1/x)$.

If we consider the application of Borel summation on the `paradigm divergent series' (PDS),~(\ref{paraDS}), :-
\begin{equation}
p(x)_{PDS}=\sum_{n=0}^{``\infty"} \Gamma(n+1) x^n,
\label{borel1}
\end{equation}
one inserts
\begin{equation}
\Gamma(n+1)=\int_0^\infty ds\,e^{-s} x^n
\label{gammaFunc}
\end{equation}
into~(\ref{borel1}), interchanges the integral and summation signs, and then does the summation to yield
\begin{equation}
p(x)_{Borel}=\int_0^\infty ds\,{e^{-s}\over 1-x s}.
\label{borel}
\end{equation}
If $x<0$, we have an alternating series, and it is asserted that the `finite value' of the divergent series~(\ref{borel1}) is
\begin{equation}
p(x)_{PDS}|_{finite}\equiv p(x)_{Borel} = -{\rm Eq. }\,(\ref{paraANS}) \,\,\,\,\,\ x < 0
\end{equation}
and yields the same results (with a minus sign) as in~(\ref{paraANS}). In this case the series is Borel summable.

If, however, $x>0$, we have a non-alternating series and there is now a pole on the line of integration at $s=1/x$ in~(\ref{borel}). A potential ambiguity now arises as one would need a prescription (e.g. principal parts) to handle the pole.  Such non-alternating series are regarded as being `non-Borel summable'. These issues do not arise when one uses the transform~(\ref{mainresult}) since it can be used in exactly the same way to evaluate non-alternating series as for alternating series (with appropriate contours for each case).

Let us now consider the convergent series representation of the exponential function $\exp(x)=\sum_{n=0}^\infty x^n/n!$. Putting $c(t)=1/\Gamma(t+1)$ in~(\ref{mainresult}), then within $C(N)$, the only poles are at $t=0,1,2,\ldots$ due to $1/(\exp(2\pi i t)-1)$, and the sum of the residues gives $\sum_{n=0}^N x^n/n!$ and since this is a convergent series, we can take $\lim_{N\rightarrow\infty} \int_{L(N)}\rightarrow 0$. However, while $1/\Gamma(t+1)$ has zeroes when $t=-1,-2,-3,\ldots$ (because $\Gamma(z)$ has poles of residue $(-1)^n/n!$ at $z=-n=0,-1,-2,\ldots$), there are no poles within $C'(-M)$ since the poles of $1/(\exp(2\pi i t)-1)$ are cancelled by the zeroes of $1/\Gamma(t+1)$. This means that $\exp(x)$ has the interesting feature that it has no divergent series representation. Therefore we find from~(\ref{mainresult})
\begin{equation}
2\pi i \sum Res(C(\infty))=\exp(x)=\sum_{n=0}^\infty {x^n\over n!} = \int_L =\int_{L'(-M)}
\end{equation}
where the last term is in fact independent of $M$.

Generally it is the combination $c(n) x^n$ that determines whether the series is convergent or divergent, e.g. the geometric series $\sum x^n$ has a finite radius of convergence -- it converges if $|x|<1$, and diverges for $x>1$. However, for many series it is primarily the nature of the coefficient function $c(n)$, rather than the power term $x^n$ that determines their convergent or divergent nature. Knowing $c(n)$ for some function, and if it gives rise to a divergent series, one can use this technique to obtain a convergent series representation of that function. It is also worth pointing out that given a convergent series representation of a function, one can use this method to obtain its divergent series representation, if such exists. 

Let us look at an example of how a divergent series arises, and then try to gain an understanding of how such objects might be handled. Consider the following integral,
\begin{equation}
Z(g)=\int_0^\infty dx e^{-x^2-g x^4}={e^{1/(8g)} \over 4\sqrt{g}}K_{1\over 4}(1/(8g))
\label{classAHO}
\end{equation}
where $g > 0$ and $K_\nu(x)$ is the modified Bessel function. If we express $\exp(-x^2)$ as a power series, interchange the integration and summation, we find that
\begin{equation}
Z(g)={1\over 4}\sum_{n=0}^\infty (-1)^n\left({1\over g}\right)^{{n\over 2}+{1\over 4}} {\Gamma\left({n\over 2}+{1\over 4}\right)\over \Gamma(n+1)}.
\label{classAHOcs}
\end{equation}
If, instead, we express $\exp(-g x^4)$ as a power series and do the same operations, we see that
\begin{equation}
``Z(g)"={1\over 2}\sum_{n=0}^{``\infty"} (-g)^n{\Gamma\left(2n+{1\over 2}\right)\over \Gamma(n+1)}.
\label{classAHOds}
\end{equation}
Using Stirling's formula $\Gamma(n+1)\approx\sqrt{2\pi n}\exp(n\ln n-n)$ for large $n$, we see that
for large enough $n$, $\Gamma(n)$ grows faster than $x^n$ for all $x$. If we take~(\ref{classAHOds}) literally we see that the series diverges as we add more and more terms, which is why quotes are used in~(\ref{classAHOds}), i.e. $``Z(g)"\ne Z(g)$. Thus the `small-g' expanson~(\ref{classAHOds}) gives a divergent series, while the `large-g' expansion~(\ref{classAHOcs}) is a convergent series. These are different series representations of the same function.

\begin{figure}[ht]
\begin{center}
\epsfig{file=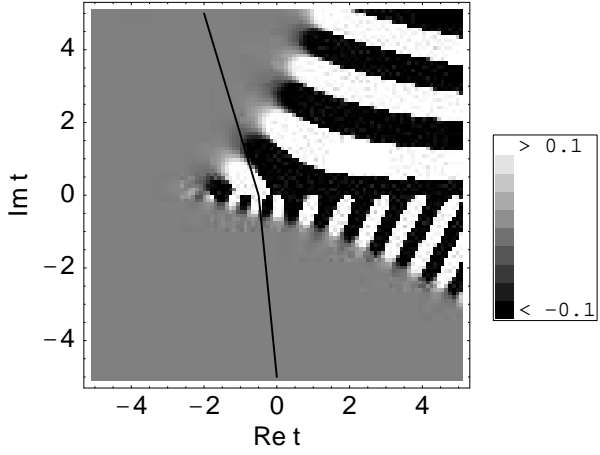,width=3.25in}
\caption{\label{zABm25}
A density plot of the real part of $((-g)^t \Gamma(2t+1/2))/(2\Gamma(t+1)(\exp(2\pi i t)-1))$
for $g=-2.5$. Light and dark bands show the oscillatory nature of the function. The dark line is a suitable contour for $L$.}
\end{center}
\end{figure}

The origin of the divergent behaviour in~(\ref{classAHOds}) is due to the fact that we have been `careless' in interchanging the order of integration and summation (of an infinite number of terms) in deriving~(\ref{classAHOds}). However, the coefficient function of the divergent series~(\ref{classAHOds})
\begin{equation}
{\Gamma\left(2n+{1\over 2}\right)\over \Gamma(n+1)}
\end{equation}
does contain {\it all} the information about the function $Z(g)$. To see this, consider the sum of a finite number of terms of~(\ref{classAHOds})
\begin{equation}
{1\over 2}\sum_{n=0}^N (-g)^n{\Gamma\left(2n+{1\over 2}\right)\over\Gamma(n+1)}=
\sum_{n=0}^N (-4g)^n{\Gamma\left(n+{1\over 4}\right)\Gamma\left(n+{3\over 4}\right)\over 2\sqrt{2\pi}\Gamma(n+1)}
\label{classAHOdsfinitesum} 
\end{equation}
where we have used $\Gamma(2n)=4^n\Gamma(n)\Gamma(n+1/2)/(2\sqrt{\pi})$ (\cite{gr} 8.335). Inserting the coefficient function into~(\ref{mainresult}) we find that the residues of the poles within $C(N)$ gives~(\ref{classAHOdsfinitesum}). It is well known that for small $g$ we can sum a few terms to give approximate (and very accurate) results for $Z(g)$ -- this is the essence of asymptotic expansions. But if we consider
\begin{eqnarray}
& &\sum_{n=0}^N (-4g)^n{\Gamma\left(n+{1\over 4}\right)\Gamma\left(n+{3\over 4}\right)\over 2\sqrt{2\pi}\Gamma(n+1)} \\
& &\hspace{5mm}-\int_{L(N)}dt\,{(-4g)^t \Gamma\left(t+{1\over 4}\right)\Gamma\left(t+{3\over 4}\right)\over 2\sqrt{2\pi}\Gamma(t+1)(\exp(2\pi i t)-1)} \nonumber
\label{classDSandINT}
\end{eqnarray}
it gives the same numerical values as the integral form of $Z(g)$ for $Re(g) > 0$. Figure~\ref{zABm25} gives a density plot of the real part of the integrand in~(\ref{classDSandINT}) for $g=-2.5$ showing the oscillatory nature, and also a suitable contour for $L$ is given by the dark line. A density plot of the imaginary part of the integrand is similar.

\begin{figure}[ht]
\begin{center}
\epsfig{file=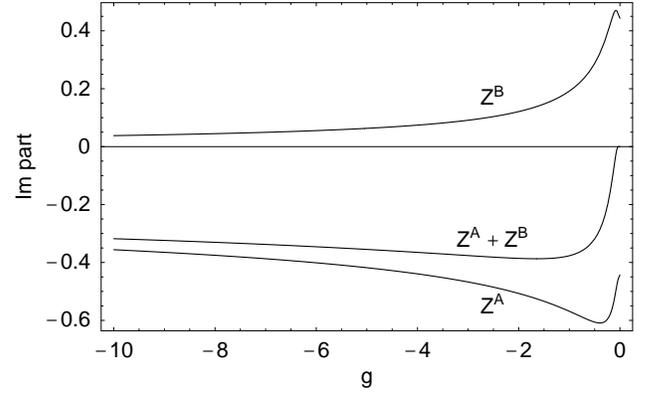,width=3.25in}
\caption{\label{zABplot}
The imaginary parts of $Z^A(g)$, $Z^B(g)$ and $Z^A(g)+Z^B(g)$ given by~(\ref{imagParts}), for $g<0$.}
\end{center}
\end{figure}

Within $C'(-M)$ there are two sets of poles, each set giving a convergent series. Letting $M\rightarrow\infty$, we find the residues of the set `A' of poles at $t=-1/4,-5/4,-9/4,\ldots$ are
\begin{equation}
-Z^A(g)=\sum_{n=0}^\infty{-\sqrt{\pi}(4 g)^{-n}\Gamma(n+1/4)\over 4 g^{1/4} \Gamma(n+1/2)\Gamma(n+1)},
\label{bess14}
\end{equation}
while the set `B' of poles of at $t=-3/4,-7/4,-11/4,\ldots$ gives residues
\begin{equation}
-Z^B(g)=\sum_{n=0}^\infty{\sqrt{\pi}(4 g)^{-n}\Gamma(n+3/4)\over 8 g^{3/4} \Gamma(n+1)\Gamma(n+3/2)}
\label{bess34}
\end{equation}
and
\begin{equation}
Z(g)=Z^A(g)+Z^B(g).
\end{equation}
If $g$ is real, the sums can be converted to Bessel functions,
\begin{eqnarray}
Z^A(g)&=&{-\pi e^{1/(8g)} \over 4\sqrt{2g}}I_{-{1\over 4}}(1/(8g)) \nonumber \\
Z^B(g)&=&{\pi e^{1/(8g)} \over 4\sqrt{2g}}I_{1\over 4}(1/(8g)).
\end{eqnarray}
For the case of $g<0$, we can express the imaginary part of $Z(g)$ as
\begin{eqnarray}
Im(Z^{A,B}(g)) &\equiv &{1\over 2}\left(Z^{A,B}(g+i \epsilon) - Z^{A,B}(g-i \epsilon)\right) \nonumber \\
Im(Z^A(g)) &=& {-\pi e^{-1/(8|g|)} \over 8|g|}I_{-{1\over 4}}(1/(8|g|)) \nonumber \\
Im(Z^B(g)) &=& {\pi e^{-{1/(8|g|)}}\over 8|g|} I_{1\over 4}(1/(8|g|)) \nonumber \\
Im(Z(g))&=&Im(Z^A(g))+Im(Z^B(g))\nonumber \\
&=& -{\pi e^{-{1/(8|g|)}}\over 4\sqrt{2 |g|}} K_{1\over 4}(1/(8|g|)).
\label{imagParts}
\end{eqnarray}
Figure~\ref{zABplot} shows a plot of the imaginary parts.

\section{One loop QED effective action}
The Euler-Heisenberg-Schwinger one-loop QED effective action~\cite{eh,schwinger,dunne}, for a constant background magnetic field $B$ can be expressed as a proper time integral :--
\begin{equation}
S=-{e^2 L^3 T B^2\over 8\pi^2} S_B(g_B)
\end{equation}
where $L^3 T$ is the space-time volume factor, and
\begin{equation}
S_B(g_B)=\int_0^\infty{ds\over s^2}\left(\coth s-{1\over s}-{s\over 3}\right)\exp\left({-s\over\sqrt{g_B}}\right)
 \label{schPT}
\end{equation}
where, $g_B=e^2 B^2/m^4$. If we expand $\coth(s)$ as a power series about $s=0$ we obtain the divergent series (and hence the use of quotes) :--
\begin{equation}
``S_B(g_B)" =16 \, g_B \sum_{n=0}^{``\infty"} c(n) g_B^n
\end{equation}
with coefficient function
\begin{eqnarray}
c(n)&=&{(-1)^{n+1} 4^n |B_{2n+4}|\over (2n+4)(2n+3)(2n+2)}\nonumber\\
&=&{-\zeta(-2n-3) (-4)^n\over 2 (2n+3)(n+1)\cos(n\pi)}
\label{schCF}
\end{eqnarray}
and $B_{2n}$ are the Bernoulli numbers, whose magnitudes are given by
\begin{equation}
|B_{2n}|={2 (2n)!\over (2\pi)^{2n}}\zeta(2n)
\end{equation}
while $\zeta(s)$ is the Riemann $\zeta$-function. We have used the following relation between the $\Gamma$- and $\zeta$-functions (\cite{gr} 9.535) :--
\begin{equation}
\Gamma(z) \zeta(z)= {(2\pi)^z \zeta(1-z)\over 2\cos(z\pi/2)}.
\label{gammazeta}
\end{equation}
Notice in~(\ref{schCF}) we have not written $1/\cos(n\pi)=(-1)^n$ for $n$ integer as we wish to maintain the analytical structure of~(\ref{gammazeta}) for complex $z$. 

If one replaces $1/\cos(n\pi)$ with $(-1)^n$ (even though they give the same divergent series), which will cancel the other $(-1)^n$ term in~(\ref{schCF}), the pole/singularity structure at $t=-1, -3/2, -2$ will give different residues from the simple (i.e. non-leading singularities) poles at $1/(t+1)$, $1/(t+3/2)$ and $1/(t+2)$. The resolution of ambiguities of this nature depends on the problem at hand. In this case, the term $1/\cos(z\pi)$ is not the same as $(-1)^z$ for complex z.

If we insert the coefficient function~(\ref{schCF}) into~(\ref{mainresult}) we find within the contour $C(N)$ there are simple poles at $t=0,1,2,\ldots$ whose residues give~(\ref{schDSnew}). Thus
\begin{equation}
S_B(g_B)={-16 g_B\over 2}\sum_{n=0}^N {\zeta(-2n-3) (-4 g_B)^n\over (2n+3)(n+1)\cos(n\pi)} -\int_{L(N)}
\label{schDSnew}
\end{equation}
 
 Within $C'(-M)$ we have simple and double poles at $t=-1,-3/2$ and $-2$; two sets of simple poles, one set at $t=-3,-4,-5,\ldots$ and another set at $t=-5/2,-7/2,-9/2,\ldots$. The residues of all poles within $C'(-M)$ are related to $S_B(g_B)$ through
\begin{equation}
S_B(g_B)=-2\pi i  \sum {\it Res(C'(-M))}+\int_{L'(-M)}.
\end{equation}
Letting $M\rightarrow\infty$, we find the closed form
\begin{eqnarray}
S_B(g_B)&=&{1\over 3}-{1\over 4 g_B}-4\zeta'\left(-1,1+{1\over 2\sqrt{g_B}}\right)
 \nonumber \\ &-&\left({1\over 6}+{1\over 4 g_B}+{1\over 2\sqrt{g_B}}\right)\log(4 g_B) 
\label{schRESULT}
\end{eqnarray}
where $\zeta'(s,a)$ is the derivative of the Hurwitz $\zeta$-function,
\begin{equation}
\zeta'(s,a)={d\over ds} \zeta(s,a).
\end{equation}
Numerical values for~(\ref{schRESULT}) can be obtained using Mathematica~\cite{math}, where $\zeta'(-1,a)$ is expressed in terms of the Barnes' G-function~\cite{adamchik}.

If instead one has constant electric fields, the one-loop effective action is
\begin{equation}
S=-{e^2 L^3 T E^2\over 8\pi^2} S_E(g_E)
\end{equation}
where $g_E=e^2 E^2/m^4$, and 
\begin{equation}
S_E(g_E)=-S_B(-g_E).
\end{equation}
One finds very good numerical agreement of 
\begin{eqnarray}
Im(S_E(g_E))&=&Im(-S_B(-g_E)) \nonumber \\
&=&{1\over \pi}\sum_{k=1}^\infty {1\over k^2}\exp\left({-\pi k\over \sqrt{g_E}}\right).
\end{eqnarray}

\section{Discussion and Conclusion}

The above examples illustrate how various series expansions of a function are related. If a problem is well posed, one would  hope to obtain well defined results. If a divergent series in $x$ arises, the starting point is to utilize the information contained in the coefficient function $c(t)$. By considering a sum of a finite number of terms, i.e. a partial sum, and expressing it as a Cauchy integral representation via~(\ref{theINT}), one then attempts to find a contour $C(N)$ that can be split into two parts $L(N)$ and $L$ as in Fig.~\ref{figure}. If a suitable contour can be found, the integral along $L(N)$ is subtracted from the partial sum, the result of which gives the integral along $L$. This in turn is related to the corresponding (convergent) series in $x^{-1}$ obtained from the same coefficient function $c(t)$.

Equation~(\ref{mainresult}), relating a series in $x$ to a series in $x^{-1}$ is an exact relation. We also see there is no essential difference between alternating (Borel summable) and non-alternating (non-Borel summable) series, in contrast to  the situation where a prescription is required to handle the potential ambiguity in the Borel summation procedure for non-alternating series (e.g. for the case of $x>0$ in~(\ref{borel})). If one has a situation where there are several $\Gamma$-functions in the numerator of $c(t)$, then the Borel summation procedure would result in each $\Gamma$-function being expressed in the integral form~(\ref{gammaFunc}), leading to a multi-dimensional integral version of~(\ref{borel}). An advantage of using~(\ref{mainresult}) is that it is a one-dimensional integral for arbitrary $c(t)$.

If there are $\Gamma$-functions in the coefficient function in the form of $\Gamma(n+\alpha)$ where $n$ is a positive integer, and $\alpha$ not an integer, then it is likely that the $\Gamma$-function reflection formula $\Gamma(x)\Gamma(1-x)=\pi/\sin(\pi x)$ will introduce phase factors in the residues of poles within $C'(-M)$. This can be seen in~(\ref{alphaCS1}) 
and~(\ref{alphaCS2}) which respectively give phases factors of the type $(-x)^m\sin(\alpha\pi)$ and 
$(-x)^{\alpha+m}\sin(\alpha\pi)$. This leads to terms in the series that will, in general, be complex if $\alpha$ is not an integer. Care should be exercised, for example, in situations where large-order perturbation results are used -- since expressing $\Gamma(3n+1/2)$, say, as $\Gamma(3n)$ for large $n$ could lead to different imaginary parts.

It is not always the case that applying~(\ref{mainresult}) to a non-alternating series will give a finite imaginary part -- for example, consider the geometric series, $\sum x^n$, for $x=2$, say, which when summed gives $1/(1-2)=-1$. 

An interesting feature of~(\ref{classAHO}) is that the divergent series representation~(\ref{classAHOds}), after the application of~(\ref{mainresult}) yields two convergent series, (\ref{bess14}) and~(\ref{bess34}), which are also valid when $g<0$, even though the integral in~(\ref{classAHO}) is undefined. Thus, in some cases, the range of applicability of the divergent (and convergent) series exceeds that of the integral representation of the function in~(\ref{classAHO}).

In general, the difficult part of a problem is to find the exact coefficient functions, and often one has to resort to some approximation. The results obtained from~(\ref{mainresult}) are only as good as our ability to find the exact coefficient function. The form of~(\ref{mainresult}) is well suited for analytical and numerical evaluation of series via contour integration (using packages like Mathematica). By expressing some divergent series as a convergent series, or even vice-versa, one might be able gain a further understanding of the nature of the solution to a problem.

\section{Acknowledgements}
I benefitted from numerous discussions with Nicole Bell, John Costella, Bob Delbourgo, Angas Hurst, Tien Kieu, Don Koks, Vic Kowalenko, John McIntosh, Ray Sawyer, Jerry Stephenson and Cameron Wellard. The facilities, and assistance of Heath O'Connell, of Information Resources Department at Fermilab is appreciated. Also I thank Gerald Dunne for valuable comments.

{}


\begin{thebibliography}{99}

\bibitem{dyson} F.~J.~Dyson, Phys.\ Rev. {\bf 85}, 631 (1952).

\bibitem{hurst} C.~A.~Hurst, Proc.\ R.\ Soc.\ London {\bf A214}, 44 (1952); Proc.\ Cambridge\ Philos.\ Soc. {\bf 48}, 625 (1952).

\bibitem{thirring} W.~Thirring, Helv.\ Phys.\ Acta {\bf 26}, 33 (1953).

\bibitem{LeGandZJ} J.~C.~Le~Guillou and J.~Zinn-Justin, Large Order Perturbation Theory; Current Physics -- Sources and Comments, Vol. 7. North Holland (Elsevier Science Publishers B.V.) 1990.

\bibitem{kleinertPhi4} H.~Kleinert and V.~Schulte-Frohlinde, Critical Properties of $\phi^4$-Theories, World Scientific, 2001.


\bibitem{weniger} E.~J.~Weniger (math.CA/0107080).

\bibitem{kowalenko} V.~Kowalenko, N.~E.~Frankel, M.~L.~Glasser and T.~Taucher, 1995, Generalised Euler-Jacobi Inversion Formula and Asymptotics Beyond All Orders (London Mathematical Society Lecture Note 214) (Cambridge: Cambridge University Press).

\bibitem{kowANDtau}  V.~Kowalenko and T.~Taucher, 1997, University of Melbourne Preprint UM-P-97/06.

\bibitem{kowANDraw}
V.~Kowalenko and A.~A.~Rawlinson,
J.\ Phys.\ A {\bf 31}, L663 (1998).

\bibitem{eh} W.~Heisenberg and H.~Euler, Z.\ Phys. {\bf 98}, 714 (1936). 

\bibitem{schwinger} J.~Schwinger, Phys.\ Rev. {\bf 82}, 664 (1951).

\bibitem{dunne}
G.~V.~Dunne and T.~M.~Hall,
Phys.\ Rev.\ D {\bf 60}, 065002 (1999)

\bibitem{gr} I.~S.~Gradshteyn and I.~M.~Ryzhik, 1980, Table of Integrals, Series and Products, Academic Press Inc. (London) Ltd.

\bibitem{math} S. Wolfram, 1999, The Mathematica Book, Wolfram Media and Cambridge University Press.

\bibitem{adamchik} V.~Adamchik,``On the Barnes function", Proceedings of the 2001 International Symposium on Symbolic and Algebraic Computation (July 22-25, 2001, London, Canada), Academic Press, 2001, pp.15-20.

\end{thebibliography}
\end{document}